\documentclass[letterpaper, 10 pt, conference]{ieeeconf}  

\IEEEoverridecommandlockouts                              
\usepackage{graphics} 
\usepackage{epsfig} 
\usepackage{amsmath} 
\usepackage{amssymb}  

\usepackage{url}
\usepackage[ruled, vlined, linesnumbered]{algorithm2e}
\usepackage{verbatim} 
\usepackage{soul, color}
\usepackage{lmodern}
\usepackage{fancyhdr}
\usepackage[utf8]{inputenc}
\usepackage{fourier} 
\usepackage{array}
\usepackage{makecell}

\SetNlSty{large}{}{:}

\makeatletter

\newcommand{\Rom}[1]{\expandafter\@slowromancap\romannumeral #1@}
\makeatother

\title{\LARGE \bf
ECG Classification System for Arrhythmia
Detection Using Convolutional Neural
Network
}

\author{Aryan Odugoudar 
\\Australian National University\\
{\tt\small aryanodugoudar143@gmail.com} \\ \\
Jaskaran Singh Walia%
\\ Vellore Institute of Technology\\
{\tt\small karanwalia2k3@gmail.com}
}

\begin{document}

\maketitle
\thispagestyle{plain}
\pagestyle{plain}

\begin{abstract}

Arrhythmia is just one of the many cardiovascular illnesses that
have been extensively studied throughout the years. Using a multi-
lead ECG data, this research describes a deep learning (DL) pipeline
technique based on a convolutional neural network (CNN)
algorithms to detect cardiovascular arrhythmia in patients. The
suggested model architecture has hidden layers with a residual
block, in addition to the input and output layers. In this study, the
classification of the ECG signals into five main groups namely: Left Bundle
Branch Block (LBBB), Right Bundle Branch Block (RBBB),
Atrial Premature Contraction (APC), Premature Ventricular
Contraction (PVC), and Normal Beat (N) are performed.
Using the MIT-BIH arrhythmia dataset, we assessed the suggested
technique. The findings show that our suggested strategy classified
15000 cases with a high accuracy of 98.2\%.

\end{abstract}

\section{INTRODUCTION}
Since cardiovascular diseases account for 10\% of all diseases and 30\% of all deaths worldwide, they constitute a challenge for global public health. 

The conventional CVD diagnosis paradigm is based on the clinical examinations and medical history of a specific patient. The patients are categorised based on these findings using a taxonomy of medical illnesses and a set of quantitative medical indicators. Unfortunately, a vast volume of heterogeneous data renders the conventional rule-based diagnosis paradigm ineffective, necessitating extensive analysis and specialised medical knowledge to attain appropriate accuracy. In poor nations where there is a shortage of medical professionals and medical equipment, the issue is more severe.

The most popular method for monitoring heart activity is the electrocardiogram (ECG), which is both accessible and non-invasive . Different types of heartbeats may typically be distinguished by meticulously evaluating ECG morphology. ECG is not typically helpful for non-stationary signals, though. The morphology of this last changes over time, and these changes can be seen in both separate cases as well as the same patient . The ability of skilled medical professionals to decipher the features of ECG signals is crucial for the early detection of arrhythmia. Doctors must have a high level of professional expertise. Consideration of computer-aided detection and diagnosis in ECG signals for cardiovascular illnesses is growing. However, it is quite challenging to create and choose the best diagnostic model with clinical implications.

As a result, a variety of ECG heartbeat recognition and classification algorithms have been created using various methods, including wavelet transform , hidden Markov models , support vector machines, and artificial neural networks. Due to the significance of accuracy in the medical field, the bulk of these ECG beat categorization techniques perform well during training but provide poor results. In this article, a convolutional neural network (CNN)-based ECG classification system based on deep learning is suggested to categorise five different types of heartbeat. Given that deep learning, particularly CNNs, have drawn a lot of attention recently for their outstanding abilities in the areas of image processing and natural signal processing, as well as their enormous potential to recognise signals.

\section{Model Analysis}

Preprocessing, heartbeat segmentation, feature extraction, and classification are the four stages of the ECG heartbeat classification process. The goal of the signal preprocessing is to remove various types of noise from the ECG signal, such as artefacts and baseline drift. In the literature, a variety of techniques for ECG signal denoising have been reported. Traditional filtering techniques include the usage of low-pass filters, Weiner filters, adaptive filters, and filter banks are included in these techniques. A lot of researchers have also worked on classification of ECG signals and feature extraction using conventional machine learning techniques. Additionally, certain statistical techniques were utilised to extract features from ECG data, including principal component analysis (PCA), the higher-order statistic (HOS) methodology, and linear discriminant analysis (LDA).

The majority of studies confirmed that wavelet transform (WT), which can concurrently extract frequency and time information, has a satisfactory outcome for ECG signal feature extraction. They classified five different beat types using the WT technique in and were 97.29\% accurate. 

Recently, the majority of classification techniques merged the two phases of feature extraction and classification by using a deep learning model. Acharya et al. [1] used nine layers of CNN to classify arrhythmia heartbeats and achieved accuracy of 94.03\% and 93.47\% with original and denoising signals, respectively. Keep in mind that the CNN model's learning capabilities will improve as the number of network layers increases. But increasing the number of network layers alone won't help with accuracy. The term for this issue is the vanishing gradients.

\begin{figure}[thpb]
      \centering
      \framebox{\parbox{2in}{\includegraphics[height=4in, width=2in]{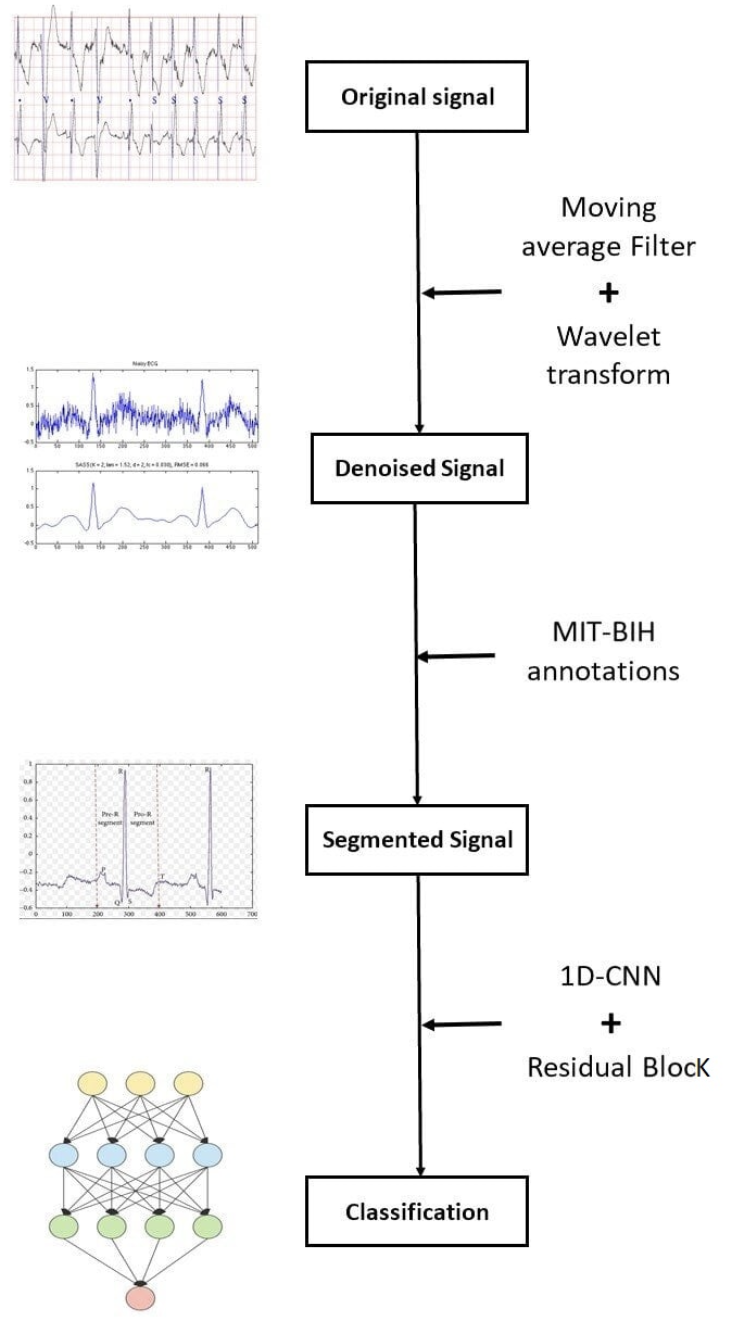}}}
      \caption{Model Pipeline}
      \label{fig:1}
\end{figure}

Normalized initialization and intermediate normalised approaches have mostly been used to resolve it [17], [18]. In fact, even with these methods, the training of deep neural networks still suffers from the previously described problem of accuracy loss as network depth increases. However, in [19], they employed a straightforward CNN model with only five layers and got a 97.5\% accuracy rate.

In this work, we start by adding two convolutional layers and using the coherent latter technique [19]. We added a residual block that contains these two convolutional layers hidden in order to identify five different types of heartbeats in order to avoid the issue of vanishing/exploding gradients. We outperformed in terms of accuracy as a result.

\section{METHODOLOGY}

\subsection{Overview of the methodology}

The recordings are initially filtered by a moving average filter and an eight-level Daubechies 4 wavelet transform in order to categorise the input ECG signal into 5 groups. Based on the MIT-BIH annotations, 200 samples are divided into each record. Then, prior to training, a dimension reduction of 180 samples is used. Finally, to achieve the feature extraction and classification of ECG signals, the processed heartbeat segments are employed directly as input data of the CNN model. Figure 1 provides a general overview of the suggested strategy.

\begin{figure}[thpb]
      \centering
      \framebox{\parbox{2in}{\includegraphics[height=2in, width=2in]{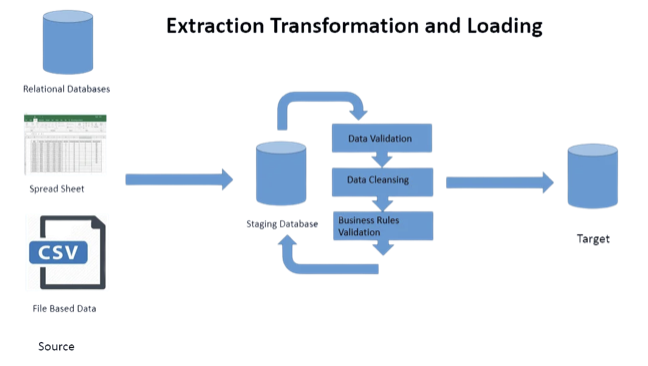}}}
      \caption{Overview of the ETL process}
      \label{fig:1}
\end{figure}

\subsection{Input data used}

PhysioNet (http://www.physionet.org) [21] hosts the MIT-BIH database, from which we pulled data for this study . 48 recordings for two-channel dynamic ECGs are present in this database. Each recording has a maximum duration of 30 minutes and a 360HZ sample frequency. Each beat is annotated by the MT-BIH to indicate the class to which it belongs. Figure I displays the number of beats per class in the Mit-Bih database. To train and test the viability of our technique, 44 ECG records from the lead II (MLII) in the database were chosen.

\begin{figure}[thpb]
      \centering
      \framebox{\parbox{2in}{\includegraphics[height=2in, width=2in]{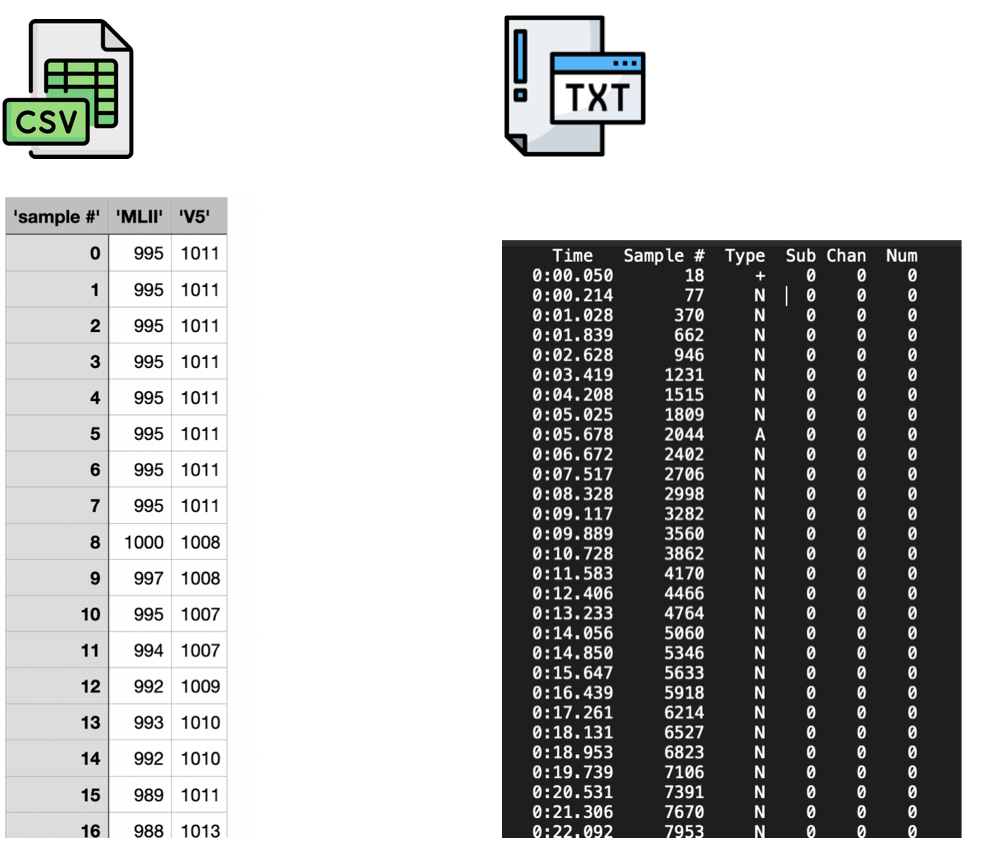}}}
      \caption{Dataset example}
      \label{fig:1}
\end{figure}

\begin{figure}[thpb]
      \centering
      \framebox{\parbox{3in}{\includegraphics[height=2in, width=3in]{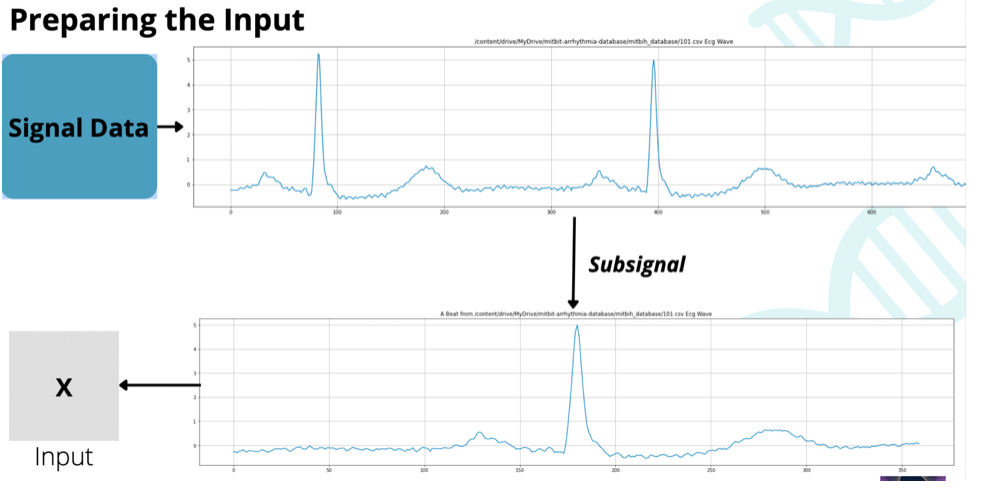}}}
      \caption{Input Signal Data}
      \label{fig:1}
\end{figure}

The 4 beats were disregarded due to their poor signal quality for post processing, as stated by the AAMI standard(102, 104, 107, 217). Our data was split into a training set (50\%) and a test set (50\%). There are 13200 non-duplicate occurrences in each collection. The total number of data chosen for each dataset is shown in Table I.

\begin{table}[h]
\caption{Classes of Arrhythmia}
\label{tab:1}
\begin{center}
\begin{tabular}{|c||c||c||c||c||c||c||c||c|}
\hline
\thead{Arrhythmia Types} & \thead{Annotations} & \thead{Total}\\
\hline
Normal Rythm(NOR) & N & 73324  \\
\hline
Left Bundle\\ Branch Block(LBBB) & L & 8071  \\
\hline
Right Bundle \\Branch Block(RBBB) & R & 6858  \\
\hline
Premature Ventricular \\Contraception(PVC) & A & 6767  \\
\hline
Atrial Premature \\Contraception(APC) & V & 1161  \\
\hline
\end{tabular}
\end{center}
\end{table}

\subsection{Data Processing}

In general, several interference noises would be easily mixed during the collection process due to the weak ECG signal and the effect of acquisition equipment. However, these sounds make it very difficult to analyse ECG readings. Therefore, prior to the classification of ECG, proper preprocessing of ECG signals is a crucial issue. Power frequency interference, baseline drift, and electromyographic interference are three common ECG signal interference sounds. Moving average filter and Daubechies 4 wavelet transform are both used to denoise the ECG signal.

In real-world scenarios, valuable signals typically appear as low-frequency or more smooth signals while noisy signals typically appear as high-frequency signals in signal processing. The high-frequency wavelet coefficients are obtained from the signals with noise when the signals are divided by the wavelet transform. Then, high-frequency wavelet coefficients are threshold processed to remove interference from power lines and electromyography. Finally, the inverse wavelet transform is used to reconstruct signals. The baseline drift noise is eliminated while moving by the average filter.

\begin{figure}[thpb]
      \centering
      \framebox{\parbox{3.2in}{\includegraphics[height=2in, width=3.2in]{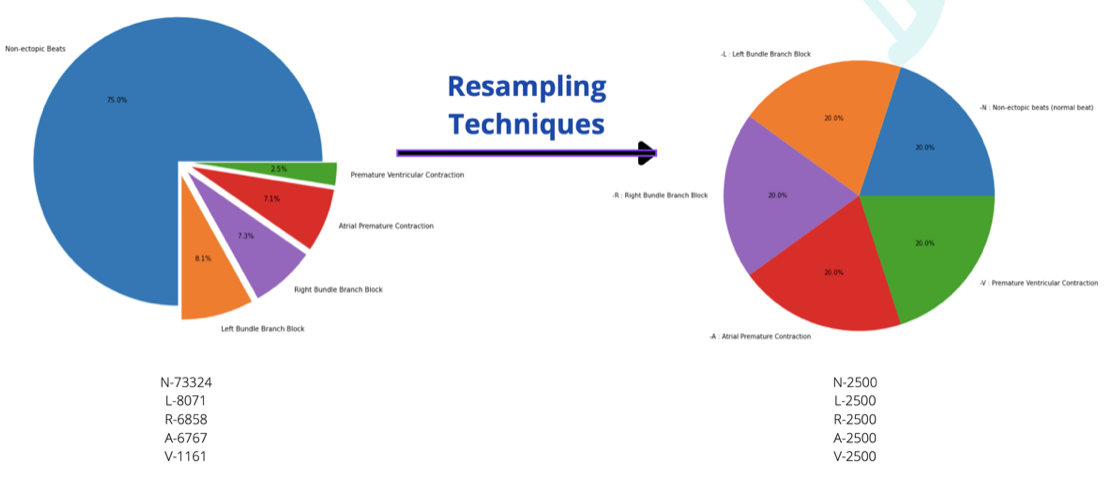}}}
      \caption{Class imbalance rebalancing}
      \label{fig:1}
\end{figure}

\subsection{CNN Model Architecture}

Different manually created features are used by traditional machine learning techniques to obtain representations of input data. Deep learning involves an autonomous learning process that progresses from low-level representations acquired over numerous layers to higher abstract representations. 

One of the most popular varieties of artificial neural networks is the CNN. A CNN is conceptually similar to a multilayer perceptron (MLP). When the network contains more than one hidden layer, an MLP becomes a deep MLP. Since each perceptron in MLP is interconnected with every other perceptron, there is a risk that the total number of parameters will increase significantly. Due to the large degree of redundancy, this is inefficient. Its disregard for spatial data is another drawback.It accepts inputs of flattened vectors. These issues were fixed by the CNN model by accounting for local connection. Additionally, all layers are somewhat loosely rather than completely attached.

\begin{figure}[thpb]
      \centering
      \framebox{\parbox{3.2in}{\includegraphics[height=2in, width=3.2in]{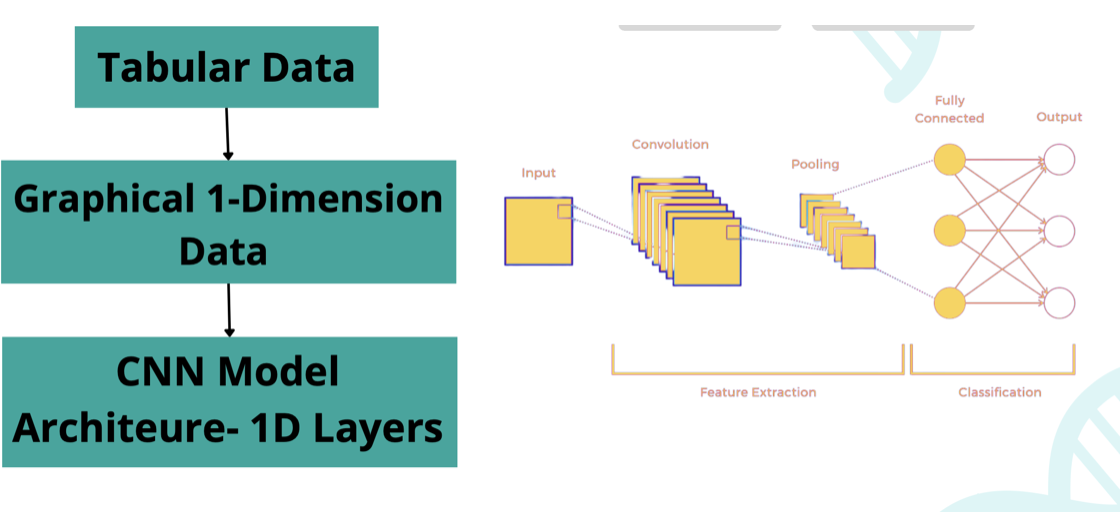}}}
      \caption{Model}
      \label{fig:1}
\end{figure}

The convolution layer, pooling layer, and fully-connected layer are the three fundamental layers of the CNN architecture. It also consists of two parts: a feature extractor that automatically learns the features from the raw input data, and a fully integrated multi-layer perception system (MLP). The convolution layer and the pooling layer are the first two layers of the feature extractor. The input is scanned in terms of dimensions by the first layer, which also utilises filters and convolution processes. The stride and filter size are two of its hyper-parameters. The output is added by a bias before being subjected to the activation function to create a feature map for the following layer. This output is known as a feature map or activation map.As the beat samples data input vector, let x0i = [X1, X2....Xn], where n is the number of samples per beat.

The output of the convolution layer is:
\[
c_{l,j} = \sigma\left(b + \sum_{j=1}^{M} w_j x_{0j}\right)
\]

where wmj is the weight for the jth feature map and mth filter index, l is the layer index, is the activation function, b is the bias term for the jth feature map, M is the kernel/filter size. The pooling layer comes just after the convolution layer. It is a process of downsampling. It helps to shrink the size of the activation map, which produces medium-level features. A layer's pooling of a feature map is determined by

			\[
P_{l,j} = \max_{r \in R} \left(c_{l,j} \cdot i \cdot T + r\right)
\]
where T is the pooling stride and R is the pooling window's size. The fully connected layer is the final layer (FC). It uses a flattened input where all neurons are coupled to each input. An activation function, a mathematical equation that defines a neural network's output, was used in each neural network. Each neuron in the network has a function connected to it that decides whether or not to activate it based on whether the input from that neuron is important to the prediction made by the model. ReLu serves as an activation function in this investigation. Considering For many different kinds of neural networks, it has evolved into the standard activation function. There isn't any difficult math.

As a result, the model can train or run faster. It is described mathematically as y = max (0,x). The appearance is similar to  A straightforward softmax classifier is utilised for beat classification and is positioned at the end of the CNN design.It is a mathematical function that turns a vector of integers into a vector of probabilities, where the probabilities of each value are inversely correlated with their relative sizes in the vector. The prediction error is determined using the loss function when the predicted output is obtained through forward propagation. Back propagation is then used to update weights by computing the gradient of the convolutional weights. The projected error propagates back on each parameter of each layer during this process. The propagation process is continued both forward and backward until a certain number of epochs are reached.

\begin{figure}[thpb]
      \centering
      \framebox{\parbox{3.2in}{\includegraphics[height=2in, width=3.2in]{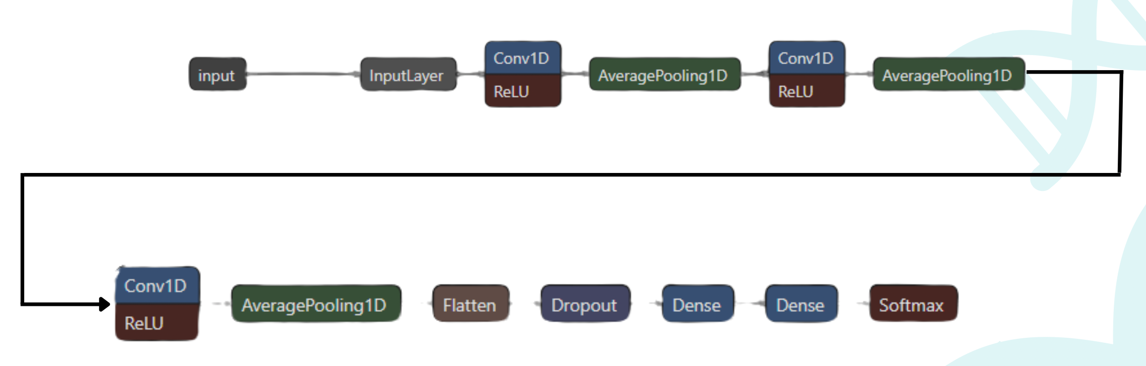}}}
      \caption{Model Pipeline Architecture}
      \label{fig:1}
\end{figure}

An important factor influencing the final classification and recognition results is the deep learning network's depth. The general concept is to make the neural network design as deep as possible. Increasing the depth will eventually hurt the deep learning network's performance though. Vanishing/exploding gradients is a problem that makes network training more challenging. The residual block was used to overcome this difficulty. It is a stack of layers configured so that each layer's output is added to a layer further down the stack. By using "shortcut connections" that skip numerous network layers, it is an enhanced deep learning algorithm for CNN that prevents these issues.

\begin{figure}[thpb]
      \centering
      \framebox{\parbox{3.2in}{\includegraphics[height=2in, width=3.2in]{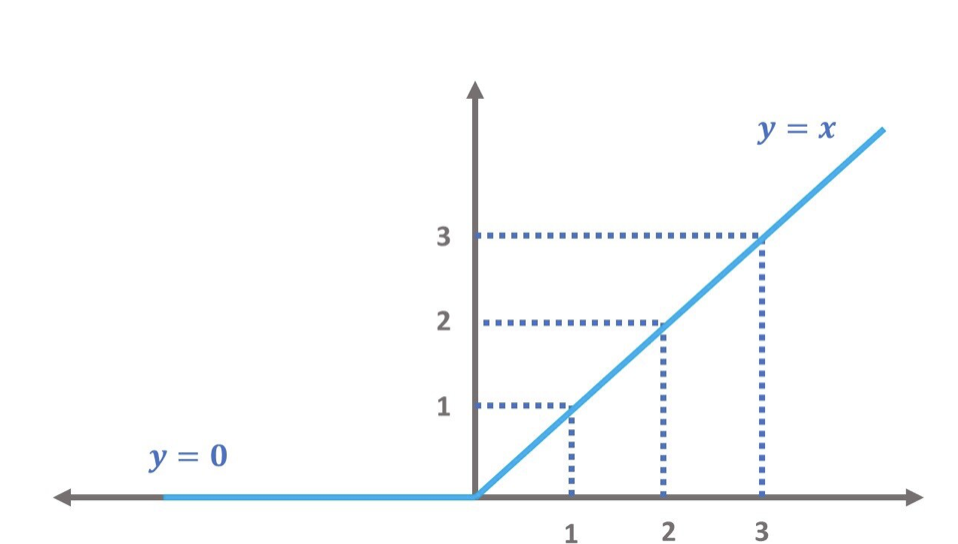}}}
      \caption{Graph for the COnvolutional Neural Networks employed}
      \label{fig:1}
\end{figure}

The two convolutional layers in the proposed model make it better than the model that was first put forth in. I added the two layers in a residual block to get around the aforementioned problem. The input layer of our suggested model architecture is composed of a segment of ECG data with 180 sample points. It has two convolutional layers, two pooling layers, two fully connected layers, and one softmax layer. The kernel sizes, strides, and number of filters on each layer are three, two, and eighteen, respectively. Reversed linear function (ReLu) was used after each convolutional layer. A residual block with two convolutional layers exists (18 convolution kernels with a length of 7 and stride 2)

\section{Experimental Results}

We used a computer with an 8GB RAM and Mac M1 processor to train our model. The international standard ECG database MIT-BIH served as the source of the experimental data. It is frequently used in ECG research and has a precise and thorough professional annotation. For training and testing purposes, this data was split into two sets, each of which has 13200 instances. 300 epochs were used for training. The dataset's batch size was 32 for each epoch, and it was expanded to include all input data. Also, 0.001 is chosen as the learning rate. Prior to training, we utilise signal rescaling to the [-1,1] range of data, which provides greater accuracy than without normalisation.

As shown by equations (3), (4), and (5), where TP stands for the true positive, TN for the true negative, FP for the false positive, and FN for the false negative, we employed the following metrics to assess the performance of our model: accuracy, specificity, and sensitivity. After experimental verification, the suggested CNN model achieved 97.8\% accuracy, 97.0\% sensitivity, and 97.32\% specificity.

\begin{figure}[thpb]
      \centering
      \framebox{\parbox{3.2in}{\includegraphics[height=2in, width=3.2in]{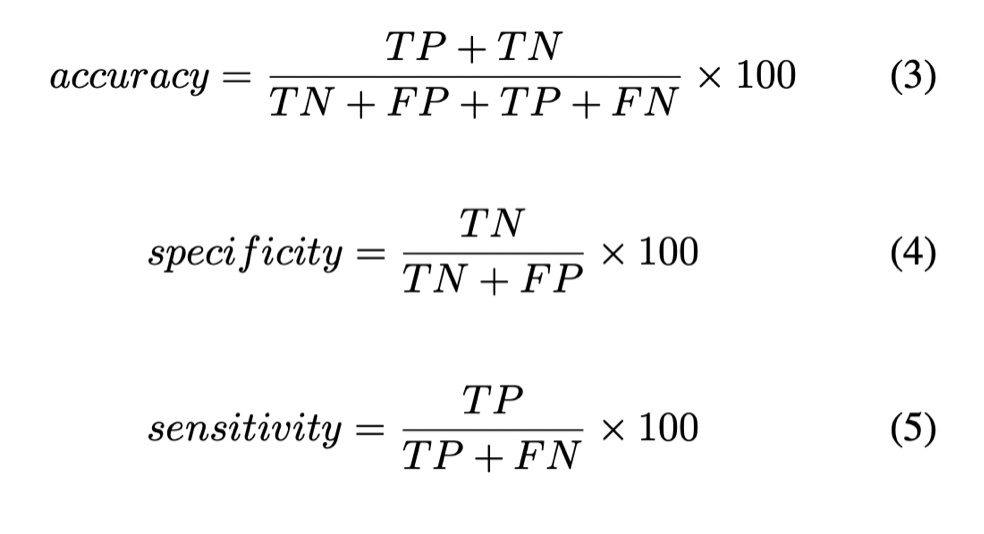}}}
      \caption{Evaluation Metrics}
      \label{fig:1}
\end{figure}

\begin{figure}[thpb]
      \centering
      \framebox{\parbox{3.2in}{\includegraphics[height=2in, width=3.2in]{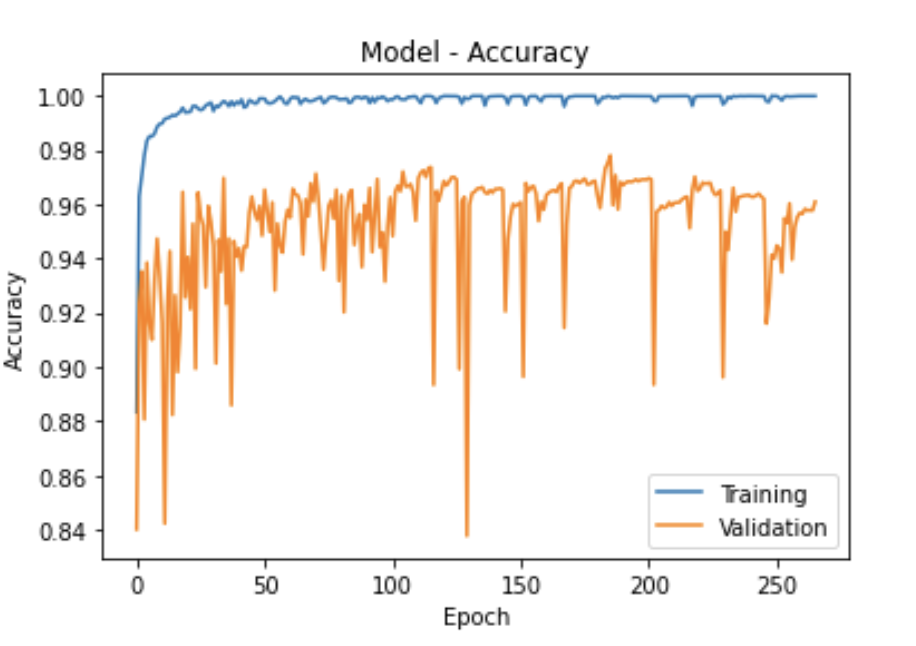}}}
      \caption{Model performance (Accuracy)}
      \label{fig:1}
\end{figure}

By employing moving average filter and wavelet transform for the preprocessing phase and 1D- CNN with residual block for classification, we can demonstrate that the suggested technique outperforms the other proposed methods in terms of ECG classification accuracy. The study's five heartbeat types are "N.L.R.A.V." Each kind corresponds to a distinct arrhythmia signal. Normal beats (N), supraventricular ectopic beats (S), ventricular ectopic beats (V), fusion beats (F), and unclassifiable beats (U) were the five forms of ECG signals according to the AAMI standard guidelines (Q).

\begin{figure}[thpb]
      \centering
      \framebox{\parbox{3.2in}{\includegraphics[height=2in, width=3.2in]{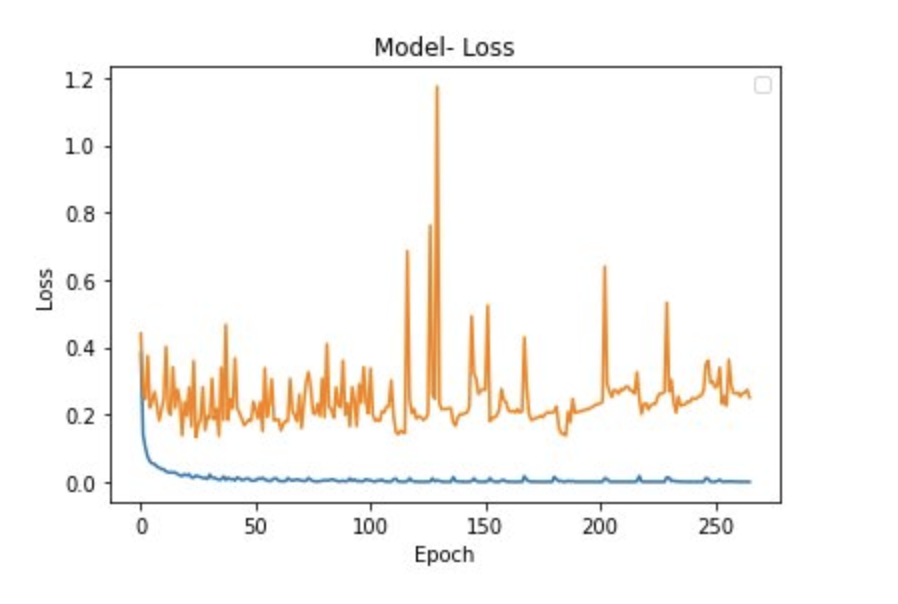}}}
      \caption{Loss Graph}
      \label{fig:1}
\end{figure}

\begin{figure}[thpb]
      \centering
      \framebox{\parbox{3.2in}{\includegraphics[height=2in, width=3.2in]{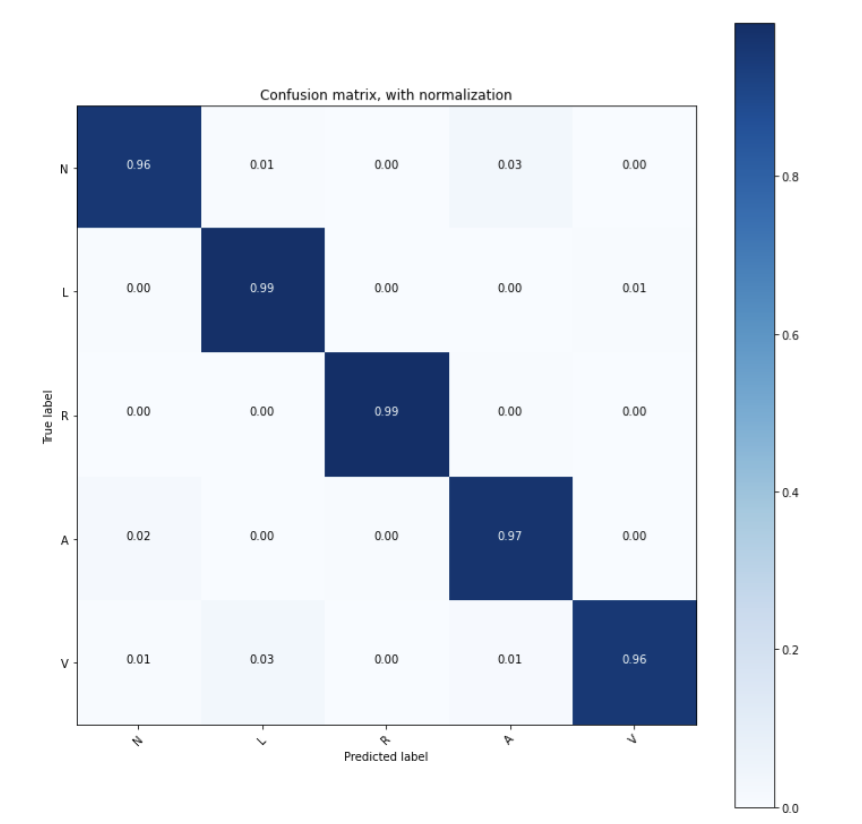}}}
      \caption{Confusion Matrix for the proposed model}
      \label{fig:1}
\end{figure}

\section{CONCLUSIONS}

The suggested model is sound and should be used in clinical settings as an auxiliary tool to assist cardiologists in identifying patients' cardiac arrhythmia. In clinical settings, this strategy will cut down on hospital ECG signal processing expenses and patient wait times. It's important to stress that a model that is highly accurate in detecting cardiovascular disease will lower medical errors. In this study, I used wavelet transform 4 in 8 levels and moving average filter to denoise the signal. The CNN model has six layers with a residual block to classify five different types of heartbeats in addition to input and output layers. I tested the trained model using the standard MIT-BIH database (lead II) in the experimental findings, and it had a 97.8\% accuracy rate. I want to increase accuracy in my future work by utilising residual architecture like ResNet or DenseNet.

\addtolength{\textheight}{-10cm}   




\end{document}